\documentclass[12pt]{article}
\makeatletter  

\@addtoreset{equation}{section} \makeatother

\begin{document}

\vspace*{0.5cm}

\newcommand{\hs}{\hspace*{0.5cm}}
\newcommand{\vs}{\vspace*{0.5cm}}
\newcommand{\be}{\begin{equation}}
\newcommand{\ee}{\end{equation}}
\newcommand{\bea}{\begin{eqnarray}}
\newcommand{\eea}{\end{eqnarray}}
\newcommand{\ben}{\begin{enumerate}}
\newcommand{\een}{\end{enumerate}}
\newcommand{\nn}{\nonumber}
\newcommand{\crn}{\nonumber \\}
\newcommand{\non}{\nonumber}
\newcommand{\noi}{\noindent}
\newcommand{\al}{\alpha}
\newcommand{\la}{\lambda}
\newcommand{\bet}{\beta}
\newcommand{\ga}{\gamma}
\newcommand{\va}{\varphi}
\newcommand{\om}{\omega}
\newcommand{\pa}{\partial}
\newcommand{\fr}{\frac}
\newcommand{\bc}{\begin{center}}
\newcommand{\ec}{\end{center}}
\newcommand{\Ga}{\Gamma}
\newcommand{\de}{\delta}
\newcommand{\De}{\Delta}
\newcommand{\ep}{\epsilon}
\newcommand{\varep}{\varepsilon}
\newcommand{\ka}{\kappa}
\newcommand{\La}{\Lambda}
\newcommand{\si}{\sigma}
\newcommand{\Si}{\Sigma}
\newcommand{\ta}{\tau}
\newcommand{\up}{\upsilon}
\newcommand{\Up}{\Upsilon}
\newcommand{\ze}{\zeta}
\newcommand{\ps}{\psi}
\newcommand{\Ps}{\Psi}
\newcommand{\ph}{\phi}
\newcommand{\vph}{\varphi}
\newcommand{\Ph}{\Phi}
\newcommand{\Om}{\Omega}
\def\lappeq{\mathrel{\rlap{\raise.5ex\hbox{$<$}}
{\lower.5ex\hbox{$\sim$}}}}

\def\sla#1{\ifmmode%
\setbox0=\hbox{$#1$}%
\setbox1=\hbox to\wd0{\hss$/$\hss}\else%
\setbox0=\hbox{#1}%
\setbox1=\hbox to\wd0{\hss/\hss}\fi%
#1\hskip-\wd0\box1 }

\bc {\Large \bf Electric Charge Quantization in\\
\vspace*{0.2cm}

$\mbox{SU}(3)_C\otimes
\mbox{SU}(3)_L \otimes \mbox{U}(1)_X$ Models}\\

\vs

{\bf Phung Van Dong}\footnote{Email: pvdong@iop.vast.ac.vn}
and {\bf  Hoang Ngoc Long}\footnote{Email: hnlong@iop.vast.ac.vn}\\

 {\it   Institute of Physics, VAST, P. O. Box 429,
Bo Ho, Hanoi 10000, Vietnam} \ec

\abstract{Basing on the general photon eigenstate and the anomaly
cancelation, we have naturally explained the electric charge
quantization in two models based on the $\mbox{SU}(3)_C\otimes
\mbox{SU}(3)_L \otimes \mbox{U}(1)_X$ gauge group, namely in the
minimal model and in the model with right-handed neutrinos. In
addition, we have  shown that the electric charges of the proton
and of the electron are opposite; and the same happens with the
neutron and  the neutrino . We argue that the electric charge
quantization is not dependent on the classical constraints on
generating mass to the fermions, but it is related closely with
the generation number problem. In fact, both problems are properly
solved as the direct consequences of the fermion content under the
anomaly free conditions.} \vs

 PACS number(s): 12.60.Cn, 11.30.Er



\section{Introduction}
\label{sec:Intro}

The $\mbox{SU}(2)_L\otimes \mbox{U}(1)_Y$ symmetry of the standard
model (SM) is a partial unification of the weak and
electromagnetic interactions. It leaves many striking features of
the physics  of our world unexplained. Some of these are the
quantization of electric charge (ECQ) and the generation number
problem (GNP). The hydrogen atom is known to be electrically
neutral to extraordinary accuracy. This implies that there is a
relation between the charges of the quarks and of the electron.
However, the electric charge operator in the SM owned with the
form $Q=T_3+Y$, while it can describe the observed charges, does
not explain them. The problem is the $Y$ generator. The values of
$T_3$ are quantized because of the non-Abelian nature of the
$\mbox{SU}(2)$ algebra. However, the values of Y are completely
arbitrary \cite{gg,chli}. They are chosen to describe the discrete
charges. However, as in the grand unified theory (GUT) \cite{vd},
both $T_3$ and $Y$ are embedded into the $\mbox{SU}(5)$ simple
group, thus the values of $Y$, like those of $T_3$ are constrained
by the structure of the algebra, hence the ECQ has been derived.
However, like the SM, the GUT also cannot explain the GNP;
moreover, this simplest version of the GUT is fairly convincingly
ruled out by the experiments on proton decay.

A very interesting alternative to explain the origin of the
generations comes from the cancelation of chiral anomalies
\cite{agg}. In particular, the models based on the
$\mbox{G}_{331}= \mbox{SU}(3)_C\otimes\mbox{SU}(3)_L \otimes
\mbox{U}(1)_X$  gauge group, also called 3-3-1 models
\cite{ppf,flt,recent}, arise as a possible solution to this
puzzle, since some of such models require three generations in
order to cancel completely chiral anomalies. In addition, in the
literature on the ECQ in some 3-3-1 models, some solutions have
been explored \cite{prs}; thus, it is hoped that the ECQ will be
solved. However, the status is still opened with many problems
which are not solved or not cleared, namely, the ECQ is obviously
not dependent on the condition to generate mass for leptons and
quarks. The usual theoretical grounds are completely unrestricted
such as the definition of the fermion content and the electric
charge operator, the relation between the ECQ and the GNP. They
are still kept as the open questions!

In this paper, we will prove that the ECQ exists in the minimal
3-3-1 model and in the 3-3-1 model with right-handed (RH)
neutrinos. We see that alternative to the GUT in which the problem
is solved on the algebra structure of the simple group; here, it
is a direct consequence from the usual fermion content in those
models. We argue that the solution for the ECQ should be based on
general laws, such as the conservation of the electric charge, the
parity invariance of the electromagnetic interaction and the
anomaly cancelation.

The rest of this paper is organized as follows: In
Sec.\ref{sec:gauge}  a brief review of the 3-3-1  models is
presented. It is emphasized that the photon eigenstate is
dependent only on form of the electric charge operator from  which
 the ECQ is derived.
 Next, in Sec.\ref{sec:refandcho}, the fermion content,
 the electric charge operator and the
classification of the models are represented. In Sec.\ref{sec:DS},
from parity invariance of the  electromagnetic vertices and
 anomaly cancelation, the ECQ is obtained. Our conclusions are
summarized in the last section - Sec.\ref{sec:conclusions}.

\section{Some remarks on the gauge sector}
\label{sec:gauge} To proceed further, in this section we review
some essential consequences on the gauge sector of any 3-3-1 model
which has been verified in \cite{ld}. Basing on these and two
general properties of the electromagnetic interaction-the
conservation of the electric charge and the parity invariance, we
get equations for the usual $\mbox{U}(1)_X$ charges. Then, the ECQ
in the 3-3-1 models is derived.

Suppose that, under the $\mbox{G}_{331}$ symmetry, there are a
fermion triplet ${\bf 3}=(f_u,f_d,f_s)^T_L$ which is composed of a
doublet $(f_u,f_d)^T_L$ and a singlet $f_{sL}$ of the
$\mbox{SU}(2)_L$ group of the SM, and an electric charge operator
(eco) in this basic owning the form \be Q=T_3+\beta
T_8+X.\label{tet}\ee To beak symmetry spontaneously, in general,
three Higgs triplets are introduced\be
\chi\sim(1,3,X_\chi),\eta\sim(1,3,X_\eta),\rho\sim(1,3,X_\rho),\ee
which must acquire the vacuum expectation values (VEVs) as
follows\bea \langle\chi\rangle^T &=&
\left(0,0,\fr{v_s}{\sqrt{2}}\right),\crn \langle\eta\rangle^T
&=& \left(\fr{v_u}{\sqrt{2}},0,0\right), \\
\langle\rho\rangle^T &=& \left(0,\fr{v_d}{\sqrt{2}},0\right).\nn
\eea The $\mbox{G}_{331}$ group is decomposed into the gauge group
of the SM by the Higgs triplet $\chi$. Next, the gauge group of
the SM  is decomposed into the
$\mbox{SU}(3)_C\otimes\mbox{U}(1)_Q$ by the two remaining Higgs
triplets $\eta, \rho$. To keep the conservation of the electric
charge, the operator Q must annihilate the vacuums: $Q
\langle\chi\rangle=0$, $Q \langle\rho\rangle=0$ and $Q
\langle\eta\rangle=0$, then we get \bea X_\eta &=& -\fr 1 2
-\fr{\beta}{2\sqrt{3}},\nn\\
X_\rho &=& \fr 1 2 -\fr{\beta}{2\sqrt{3}}, \label{eqc}\\ X_\chi
&=& \fr{\beta}{\sqrt{3}},\nn\eea which are the fixing conditions
for the $\mbox{U}(1)_X$ charges of the Higgs scalars. They yield
\be X_\eta+X_\rho+X_\chi=0.\ee

The mass Lagrangian for the neutral gauge bosons is given by
\cite{ld}\be {\cal L}_{mass} = \fr 1 2 V^T M^2 V, \ee where
$V^T=(W^3,W^8,B)$ and \be
M^2=\fr 1 4 g^2\left(%
\begin{array}{ccc}
  m_{11} & m_{12} & m_{13} \\
  m_{12} & m_{22} & m_{23} \\
  m_{13} & m_{23} & m_{33} \\
\end{array}%
\right), \ee with \bea m_{11}&=&v^2_u+v^2_d,\crn
m_{12}&=&\fr{1}{\sqrt{3}}\left(v^2_u-v^2_d\right),\crn
m_{13}&=&\fr{t}{\sqrt{6}}\left[v^2_u\left(-1-\fr{\beta}{
\sqrt{3}}\right)-v^2_d\left(1-\fr{\beta}{\sqrt{3}}\right)\right],\crn
m_{22}&=&\fr{1}{3}\left(v^2_u+v^2_d+4v^2_s\right),\crn
m_{23}&=&\fr{t}{3\sqrt{2}}\left[v^2_u\left(-1-\fr{\beta}{
\sqrt{3}}\right)+v^2_d\left(1-\fr{\beta}{\sqrt{3}}
\right)-v^2_s\fr{4\beta}{\sqrt{3}}\right],\crn
m_{33}&=&\fr{t^2}{6}\left[v^2_u\left(-1-\fr{\beta}{\sqrt{3}}
\right)^2+v^2_d\left(1-\fr{\beta}{\sqrt{3}}\right)^2+v^2_s\left(
\fr{2\beta}{\sqrt{3}}\right)^2\right].\nn \eea Here $t\equiv
g_X/g$ with $g_X,$ $g$ are the gauge coupling constants of the
$\mbox{U}(1)_X$ and $\mbox{SU}(3)_L$ groups, respectively.  We
have shown \cite{ld}, for any 3-3-1 model (containing Higgs
triplets, antitriplets as well as sextets or any necessary Higgs
scalar), the mass matrix of the neutral gauge bosons always has
the above form. In addition, if additional Higgs scalars have
non-zero VEVs, one just makes the following appropriate replaces
   \bea v^2_u &\rightarrow& v^2_u+v^2_{u1}+v^2_{u2}+...,\nn\\
   v^2_d &\rightarrow & v^2_d+v^2_{d1}+v^2_{d2}+...,\nn\\
   v^2_s &\rightarrow & v^2_s+v^2_{s1}+v^2_{s2}+...,\nn\eea
where $v_{ui},v_{dj},v_{sk}$ are the VEVs of the neutral members
in the additional Higgs, respectively. Thus, this is the general
form of the mass matrix for the neutral gauge boson sector.

It can be checked that the matrix $M^2$ has a {\it non-degenerate}
zero eigenvalue. Therefore, the zero eigenvalue is identified with
the photon mass, $M^2_\ga=0$. The physical photon field $A_\mu$ is
directly defined from the equation $M^2A_\mu=0$: \be
A_\mu=\fr{t}{\sqrt{6+(1+\beta^2)t^2}}W^3_\mu+\fr{\beta
t}{\sqrt{6+(1+\beta^2)t^2}}W^8_\mu+
\fr{\sqrt{6}}{\sqrt{6+(1+\beta^2)t^2}}B_\mu.\label{ga}\ee Hence,
for any 3-3-1 model, the photon eigenstate and mass are {\it
independent} on the VEVs structure. These are a natural
consequence of the $\mbox{U}(1)_Q$ invariance - the conservation
of the electric charge. Moreover, to be consistent with the QED
based on the unbroken $\mbox{U}(1)_Q$ gauge group, the photon
field has to keep {\it the general properties of the
electromagnetic interaction in the framework of the 3-3-1 model,
such as the parity invariant nature} \cite{qkpy} (for more
discussions, see \cite{parity1}). These would help us to obtain
some consequences related to quantities which are independent on
VEVs structure such as the matching of gauge coupling constants
\cite{ld}\footnote{As a result, the condition for matching of the
gauge coupling constants in any 3-3-1 model is very natural as
done in the SM.} and the ECQ.

Next, using (\ref{ga}) we write the coupling of the up member with
the photon, $\bar{f}_uf_u\ga$ (also see \cite{ld}): \bea {\cal
L}^{em}_{\bar{f}_u f_u\ga} &=& \bar{f}_{uL} i \ga^\mu
\left[\fr{ig}{2}\fr{t}{\sqrt{6+(1+\beta^2)t^2}}+\fr{ig}{2\sqrt{3}}\fr{\beta
t}{\sqrt{6+(1+\beta^2)t^2}}\right.\crn &+&\left.
\fr{ig_X}{\sqrt{6}}X_{3}\fr{\sqrt{6}}{\sqrt{6+(1+\beta^2)t^2}}
\right]A_\mu f_{uL}\nn\\&+&\bar{f}_{uR} i \ga^\mu
\left[\fr{ig_X}{\sqrt{6}}X_{f_u
}\fr{\sqrt{6}}{\sqrt{6+(1+\beta^2)t^2}} \right] A_\mu f_{uR} \nn
\\ &=&-\fr{(X_{3}-X_\eta)g_X }{\sqrt{6+(1+\beta^2)t^2}}
\bar{f}_{uL} \ga^\mu f_{uL}A_\mu-\fr{X_{f_u}g_X
}{\sqrt{6+(1+\beta^2)t^2}}\bar{f}_{uR} \ga^\mu f_{uR} A_\mu, \nn
\eea where $X_3$ and $X_{f_u}$ are the $\mbox{U}(1)_X$ charges of
the {\bf 3} triplet and of the $f_{uR}$ singlet, respectively.
Since the electromagnetic interaction is invariant under the
parity transformation \cite{qkpy}, then we get\be
X_{f_u}=X_3-X_\eta.\label{e1}\ee Similarly for the vertices
$\bar{f}_df_d\ga$
and $\bar{f}_sf_s\ga$, we get \bea X_{f_d}&=&X_3-X_\rho,\label{e2}\\
X_{f_s}&=&X_3-X_\chi,\label{e3}\eea where $X_{f_d}$, $X_{f_s}$ are
the $\mbox{U}(1)_X$ charges of the $\mbox{SU}(3)_L$ singlets
$f_{dR}$ and $f_{sR}$, respectively. Note that $f_{uR}$, $f_{dR}$
and $f_{sR}$ are the right-handed counterparts of the triplet
$(f_u,f_d,f_s)^T_L$.

For a fermion antitriplet ${\bf 3^*}=(f'_d,-f'_u,f'_s)^T_L$ which
is composed of an antidoublet $(f'_d,-f'_u)^T_L$ and a singlet
$f'_{sL}$ of the $\mbox{SU}(2)_L$ group of the SM with $f'_{uR}$,
$f'_{dR}$ and $f'_{sR}$ are its right-handed counterparts, we also
have \bea
X_{f'_d}&=&X_{3^*}+X_\eta,\label{en1}\\
X_{f'_u}&=&X_{3^*}+X_\rho,\label{en2}\\
X_{f'_s}&=&X_{3^*}+X_\chi.\label{en3}\eea Here $X_{3^*}$,
$X_{f'_u}$, $X_{f'_d}$ and $X_{f'_s}$ stand for the
$\mbox{U}(1)_X$ charges of the antitriplet and singlets $f'_{uR}$,
$f'_{dR}$ and $f'_{sR}$, respectively.

As we will see, the lepton sector owns the triplets either ${\bf
3_l}=(\nu_l,l,\nu^C_l)^T_L$ or ${\bf 3_l}=(\nu_l,l,l^C)^T_L$,
where $\nu^C_{lL}=(\nu_{lR})^C$, $l^C_L=(l_R)^C$ with $C$ is the
conjugate operator and $l$ stands for the lepton $e$, $\mu$,
$\tau$. Then, the two equations either (\ref{e1}) and (\ref{e3})
or (\ref{e2}) and (\ref{e3}) are replaced by equation one
either\be X_{3_l}=\fr{X_\eta+X_\chi}{2},\label{f1}\ee or \be
X_{3_l}=\fr{X_\rho+X_\chi}{2}, \hs l = e, \mu, \tau,\label{f2}\ee
respectively, which  is directly obtained from the vertex
$\bar{\nu_{l}}\nu_l\ga$ or $\bar{l}l\ga$. It is worth to mention
on significance of the equation (\ref{f1}) or (\ref{f2}) which is
{\it the fixing condition for the $\mbox{U}(1)_X$ charge of the
lepton triplet as a natural consequence of the lepton content}.
Further, with anomaly cancelation, the charges for all remaining
chiral fermions are also fixed. Hence, these give the constraints
on the hypercharge values $Y=\beta T_8+X$ from which the ECQ in
the 3-3-1 models will explicitly be   explained. Otherwise, if
(\ref{f1}), (\ref{f2}) do not exist in some lepton content, there
is not the ECQ unless add auxiliary conditions \cite{prs} such as
on Majorana neutrino  mass, non-RH neutrino singlets, etc. Thus,
this means that (\ref{f1}) or (\ref{f2}) is {\it quantized
condition}.

It is to be emphasized that if the ECQ in the GUT has been found
by the mean of the algebra structure of the simple group, here for
the 3-3-1 models, by their fermion content (the fermion structure
under the anomaly free conditions).

\section{Fermion content}
\label{sec:refandcho}

In the framework of the 3-3-1 models, the essential basic concepts
for building the models such as the fermion representations
(reps), the electric charge operator, the anomaly cancelation and
the fermion content will be explained. However, for our purpose in
studying the ECQ, it is necessary to note that the electric
charges of the particles will be kept as parameters.

The SM is well done with the $\mbox{SU}(2)_L$ doublets for the
left-handed chiral spinors and the $\mbox{SU}(2)_L$ singlets for
the right-handed chiral spinors. Each generation of the SM
consists of \bea \mbox{the doublets: }&&(\nu_{l}, l)^T_L, \mbox{ }
(u^\al, d^\al)^T_L \crn \mbox{and the singlets:  }&&\nu_{lR},
\mbox{  }l_R, \mbox{ }u^\al_R, \mbox{  }d^\al_R,\label{coun} \eea
where $L, R$ stand for the left-handed and the right-handed
counterparts, respectively, $\al$ is the color index. Here $l=e,
\mbox{  }\mu, ...$; $\nu_l=\nu_e, \mbox{ }\nu_\mu, ...$; $u=u,
\mbox{ }c, ...$ and $d=d, \mbox{  }s, ...$ are the lepton and the
quark particles in each generation, respectively.

Under the gauge symmetry $\mbox{G}_{331}$, the fermions transform
like triplets {\bf 3}, antitriplets {\bf 3$^*$} or singlets {\bf
1} of the $\mbox{SU}(3)_L$ group. Requiring the models at low
energy to be fitly with the SM, the $\mbox{G}_{331}$ symmetry must
be spontaneously broken down that of the SM. Thus, the triplets or
antitriplets are composed of the doublets {\bf 2} or antitriplets
{\bf 2$^*$} and singlets {\bf 1} of the $\mbox{SU}(2)_L$ group of
the SM. The decomposional rule into the SM for the triplets yields
\bea
(\nu_l,l,S^{l})^T_L &=& (\nu_l,l)^T_L\oplus S^l_{L},\\
(u,d,S^{q})^T_L&=&(u,d)^T_L\oplus S^{q}_L.\eea Similarly for the
antitriplets\bea
(l,-\nu_l,S'^{l})^T_L &=& (l,-\nu_l)^T_L\oplus S'^{l}_L,\\
(d,-u,S'^{q})^T_L&=&(d,-u)^T_L\oplus S'^{q}_L, \eea where $S^{l}$,
$S'^{l}$, $S^q$, $S'^q$ stand for the lepton and quark singlets,
respectively. Note that if $(f_u,f_d)^T_L$ is a doublet of the
$\mbox{SU}(2)_L$, then $(f_d,-f_u)^T_L$ is its antidoublet, they
are equivalent and real reps.

Since the right-handed leptons in (\ref{coun}) are color singlets,
they are put in the singlet $S^l$ or $S'^l$ by two ways
\cite{ppf,flt}:\bea (\nu_{lR})^C=S^l_L, \mbox{ or }
S'^l_L\label{hl}\eea and \bea (l_R)^C=S^l_L, \mbox{ or }
S'^l_L.\label{pf}\eea However, under the $\mbox{G}_{331}$ and the
Lorentz invariance, we cannot put the left-handed antiquarks in
the bottom (singlet) of the triplets, so the existence of the
exotic quarks is not able to avoid in all 3-3-1 models. In
addition, the exotic leptons can be also in the singlets (in
bottom of the lepton triplets or antitriplets), however, they are
not considered here.

The fermion content under the $\mbox{G}_{331}$ symmetry must be
satisfied with the following criteria: \ben
\item All singlets of the lepton triplets and antitriplets are
either $(\nu_{lR})^C$ or $(l_R)^C$.
\item Both $(\nu_{l},l,\nu^C _l) ^T _L$, $(l',-\nu_{l'},\nu^C _{l'})^T _L$ and
$(\nu_{l},l,l^C)^T_L$, $(l',-\nu_{l'},l'^C)^T_L$ are not conjugate
pairs, namely triplet and antitriplet.\een Without loss of
generality due to the second criterion, we can put the left-handed
leptons in the triplets. Hence, on the first criterion there are
two models: with $(l_R)^C$ called the minimal 3-3-1 model
\cite{ppf}, and also for $(\nu_{lR})^C$ called the 3-3-1 model
with RH neutrinos \cite{flt}. In this paper, they are called {\it
usual} 3-3-1 models.

Due to the conservation and additive nature of the electric
charge, the eco must be embedded in the neutral generators of the
$\mbox{SU}(3)_L\otimes\mbox{U}(1)_X$ group: \be Q =\al T_3+\beta
T_8+\ga X.\label{Q}\ee Here the $\mbox{SU}(3)_L$ charges
$T_3=\la_3/2$, $T_8=\la_8/2$ with $\la_3$, $\la_8$ are the two
diagonal Gell-Mann matrices, and $X$ is the $\mbox{U}(1)_X$
charge. Without loss of generality, the $\ga$ coefficient can be
normalized to $1$ due to a scaling symmetry, $g_X\rightarrow\ga
g_X$, $X\rightarrow X/\ga$, where $g_X$ is the $\mbox{U}(1)_X$
coupling constant \cite{foot}. Finally, the two remaining
coefficients $\al$ and $\beta$ get the same dimension of the
electric charge. At the breaking point, the $\mbox{SU}(3)_L$ group
is embedded properly in the $\mbox{SU}(2)_L$ group of the SM,
therefore, the gauge boson $W$ takes an electric charged value
equal $+\al$ or $-\al$. To see this, we should apply the eco
(\ref{Q}) on a $\mbox{SU}(3)_L$ triplet, ${\bf
3}=(f_u,f_d,f_s)^T_L$, \bea \fr 1 2 \al
+\fr{1}{2\sqrt{3}}\beta+\ga X_{\bf 3} &=& q_{f_u},\crn -\fr 1 2
\al
+\fr{1}{2\sqrt{3}}\beta+\ga X_{\bf 3} &=& q_{f_d}, \label{he}\\
-\fr{1}{\sqrt{3}}\beta+\ga X_{\bf 3} &=& q_{f_s},\nn \eea with
$q_{f_u}$, $q_{f_d}$ and $q_{f_s}$ are the electric charges of the
members. Hence, the electric charge of W is $q_{f_u}-q_{f_d}=\al$.
The normalization of the eco is undetermined, however we can
always use the freedom in assigning the scale of the electric
charge by putting the charged $W$ in unit, $\al=1$ \cite{foot}.
The eco is given by\be Q =T_3+\beta T_8+X.\label{Q2} \ee

When the $Y$ hypercharge is embedded into the
$\mbox{SU}(3)_L\otimes\mbox{U}(1)_X$ group, it is a linear
combination of two terms, $Y=\beta T_8+X$. The first term in the
$\mbox{SU}(3)_L$ is constrained by the structure of the algebra
therefore quantized \cite{gg}; and, the second term in the
$\mbox{U}(1)_X$ with the values are kept as undetermined
parameters. This differs the GUT from enlarging to the simple
group, hence the ECQ is a direct consequence. However, as in the
previous section, since  all X charges are fixed, the present ECQ
signs a different structure which refers to the particle reps.

Noting that the presence of the coefficient $\beta$ signifies that
the first term of the $Y$ hypercharge is not properly normalized
to be one of the $\mbox{SU}(3)_L$ generators which have their
scale fixed by the non-linear commutation relations \cite{chli},
\bea \left[T_a,T_b \right]=if_{abc}T_c,\nn\eea with \bea {\mbox
Tr}[T_aT_b]=\delta_{ab}/2.\nn\eea The value of $\beta$ is obtained
by comparing in the fundamental rep the values of $T_8$ and the
hypercharge values of the particles in some  multiplet.

The action of the eco on an antitriplet ${\bf
3^*}=(f'_d,-f'_u,f'_s)^T$ is thanked to the usual rule\bea Q{\bf
3}=q_3{\bf 3},\eea which yields \be Q{\bf 3^*}=q_3{\bf
3^*}=-q_{3^*}{\bf 3^*}.\label{g}\ee Noting on the minus sign in
the r.h.s of (\ref{g}), we get $X_{3^*}=-X_3=-\mbox{Tr}Q$.

Demanding for the fermion $\mbox{SU}(3)_C$ reps to be vector-like
and the color number $N_C=3$, we get the non-trivial triangular
anomaly cancelation conditions \cite{rdm1,rdm2} as follows\bea
\left[\mbox{SU}(3)_C\right]^2\otimes \mbox{U}(1)_X&:&
3X^L_q-{\sum_{singlet}X^R_q}=0,\label{a1}\\
\left[\mbox{SU}(3)_L\right]^3&:&\fr 1 2 A_{\al\beta\ga}=0,\label{a2}\\
\left[\mbox{SU}(3)_L\right]^2\otimes
\mbox{U}(1)_X&:&\sum_{family}X^L_l+3\sum_{family}X^L_q =0,\label{a3}\\
\left[Grav\right]^2\otimes
\mbox{U}(1)_X&:&3\sum_{family}X^L_l+9\sum_{family}X^L_q\crn &-&3
\sum_{family}\sum_{singlet}X^R_q-\sum_{family}\sum_{singlet}X^R_l=0,\label{a4}\\
\left[\mbox{U}(1)_X\right]^3&:&3\sum_{family}(X^L_l)^3+
9\sum_{family}(X^L_q)^3\crn &-&
3\sum_{family}\sum_{singlet}(X^R_q)^3-\sum_{family}\sum_{singlet}(X^R_l)^3=0.
\label{a5}\eea Here $X^L_l$, $X^L_q$, $X^R_l$ and $X^R_q$ refer to
the $\mbox{U}(1)_X$ charges of the left-handed lepton, quark
triplets or antitriplets and the right-handed lepton, quark
singlets, respectively.

The cancelation of the $[\mbox{SU}(3)_L]^3$ anomaly (\ref{a2})
demands for the number of fermion triplets to be the same as that
of antitriplets. As mentioned above the $\mbox{SU}(2)_L$ doublet
and antidoublet are equivalent and real, hence,  all the
left-handed leptons and quarks are always ordered in the doublets.
Moreover, using two conditions such as some known fermion
generations are completely free from anomaly and the anomaly over
all the quark and lepton generations must be canceled, we deduce
that the number of the quark generations must be equal to that of
the leptons. So, if $N_f$ is the number of the fermion
generations; thus, it is also the number of the lepton triplets as
mentioned above. And, k is the number of the quark generations
which are ordered in the triplets; then, there are the remaining
$N_f-k$ quark generations therefore in the antitriplets,
satisfying \bea N_f+3k=3(N_f-k) &\Rightarrow& N_f=3k.\eea Hence,
the generation number $N_f$ is a multiple of three. If further,
one adds the condition of the QCD asymptotic freedom, which is
valid only when the quark generation number is to be less than
five, then it follows that $N_f$ is equal to $3$, and hence $k=1$.

Therefore, the classification of the 3-3-1 models is given as
follows, \ben \item The 3-3-1 model with RH neutrinos which the
fermion reps are ordered by \bea
\left(\nu_{l},l_,\nu^C_l\right)^T_L&\sim&(1,3,X^L_l),
l=e,\mu,\tau,\\
l_{R}&\sim&(1,1,X^R_{l}),\\
\left(u_{3L},d_{3L},s_{3L}\right)&\sim&(3,3,X^L_{q_3}),\\
\left(u_{iL},d_{iL},s_{iL}\right)&\sim&(3,3^*,X^L_{q_i}),i=1,2,\\
u_{iR},u_{3R}&\sim&(3,1,X^R_{u_i}), (3,1,X^R_{u_3}), \mbox{ also
for } d \mbox{ and } s.\eea \item The minimal 3-3-1 model with the
fermion reps read \bea
\left(\nu_{l},l,l^C\right)^T_L&\sim&(1,3,X^L_l),
l=e,\mu,\tau,\\
\nu_{lR}&\sim&(1,1,X^R_{\nu_l}),\\
\left(u_{3L},d_{3L},s_{3L}\right)&\sim&(3,3,X^L_{q_3}),\\
\left(u_{iL},d_{iL},s_{iL}\right)&\sim&(3,3^*,X^L_{q_i}),i=1,2,\\
u_{iR},u_{3R}&\sim&(3,1,X^R_{u_i}), (3,1,X^R_{u_3}), \mbox{ also
for } d \mbox{ and } s.\eea \een Here, the (c,f,X) denotes the
respective quantum numbers to the color, the flavor and the
X-charge, and $s_a$, $a=1,2,3$ are the added exotic quarks.

\section{The ECQ}

\label{sec:DS} Now  we turn on the ECQ in the 3-3-1 models. We
first deal with the minimal version.

\subsection{The ECQ in  the  minimal   model}
It is known that, the electromagnetic interaction is invariant
under parity transformation. Using this property and  anomaly
cancelation we will get the  needed ECQ. Let us deal with the
lepton sector.
\subsubsection{The ECQ in the lepton sector} \label{subsec:P}
For the minimal model with the given lepton triplets, using  Eq.
(\ref{f2}), we get
\bea X^L_{l}&=&\fr{X_\rho+X_\chi}{2}\nn\\
&=&-\fr{X_\eta}{2},\mbox{
 }l=e,\mu,\tau.\label{eqe}\eea Applying
Eq. (\ref{e1}) for the neutrinos, we have \be
X^R_{\nu_l}=X^L_l-X_\eta=-\fr{3}{2}X_\eta,\mbox{
 } l=e,\mu,\tau.\ee
Therefore, the application of the eco on the lepton triplets and
the neutrino singlets yields the electric charges for the leptons
as follows \bea q_{\nu_l}&=&\fr{3+\sqrt{3}\beta}{4},\\
q_{l}&=&\fr{-1+\sqrt{3}\beta}{4},\mbox{
 }l=e,\mu,\tau.\label{kq1}\eea

So, with the help of the parity invariance of the electromagnetic
vertices for all leptons, the ECQ of the lepton sector is derived.
The electric charges of all leptons in the model are defined in
terms of the $\beta$.

\subsubsection{The ECQ in the quark sector} \label{subsec:FQN}
Applying the equations (\ref{e1}), (\ref{e2}) and (\ref{e3}) for
the quark triplets; and, (\ref{en1}), (\ref{en2}) and (\ref{en3})
for the quark antitriplets, we get \bea
X^R_{u_3}&=&X^L_{q_3}-X_\eta,\mbox{  }
X^R_{u_i}=X^L_{q_i}+X_\rho,\label{eq10}\\
X^R_{d_3}&=&X^L_{q_3}-X_\rho,\mbox{  }
X^R_{d_i}=X^L_{q_i}+X_\eta,\label{eq13}\\
X^R_{s_3}&=&X^L_{q_3}-X_\chi,\mbox{  }
X^R_{s_i}=X^L_{q_i}+X_\chi.\label{eq14} \eea

With these equations, we can write all Yukawa couplings to
generate mass to all quarks \bea {\cal L}_Y&=&h^s_{33}\bar{q}_3
s_3 \chi + h^s_{ii}\bar{q_i} s_i \chi^*\crn &+&h^u_{33}\bar{q}_3
u_3 \eta + h^u_{ii}\bar{q_i} u_i \rho^*\crn &+&h^d_{33}\bar{q}_3
d_3 \rho + h^d_{ii}\bar{q_i} d_i \eta^*+ h.c..\label{ly}\eea

Since the CKM matrix is non-diagonal, there are the flavor mixing
terms in the Lagrangian (\ref{ly}). Therefore, the some terms in
(\ref{ly}) must be changed as follows \cite{foot}\bea \bar{q}_3
u_3 \eta &\rightarrow& \bar{q}_3 u_a\eta, \mbox{ }a=1,\mbox{
}2,\mbox{ }3,\crn \bar{q}_3 d_3 \rho &\rightarrow& \bar{q}_3
d_a\rho,\crn \bar{q_i} u_i \rho^*&\rightarrow &\bar{q_i} u_a
\rho^*,\mbox{ }i=1,\mbox{ }2,\crn \bar{q_i} d_i \eta^*&
\rightarrow &\bar{q_i} d_a \eta^*.\eea Under the $\mbox{U}(1)_Q$
invariance, we have \bea X^R_{u_1}&=&X^R_{u_2}=X^R_{u_3}\equiv
X^R_u,\crn X^R_{d_1}&=&X^R_{d_2}=X^R_{d_3}\equiv
X^R_d.\label{eqd}\eea Thus, it is easy to get \bea X^L_{q_1}&=&
X^L_{q_2}\equiv X^L_{q},\crn X^R_{s_1}&=& X^R_{s_2}\equiv
X^R_{s}.\nn\eea

Using the anomaly cancelation (\ref{a3}), we have \bea
2X^L_{q}+X^L_{q_3}&=&\fr{1}{2}X_\eta.\label{eq11}\eea Combination
of (\ref{eq13}) and (\ref{eqd}) yields \be
X^L_q+X_\eta=X^L_{q_3}-X_\rho.\label{eq22}\ee
From (\ref{eq11}) and (\ref{eq22}) it follows
 \bea X^L_q&=&\fr 1 6 (X_\eta+2X_\chi),\label{eq49}\\
X^L_{q_3}&=&\fr 1 6 (X_\eta-4X_\chi).\label{eq50}\eea With the
help of (\ref{eq49}) and (\ref{eq50}), we can express  all charges
for all right-handed quark counterparts in terms of  $X_\eta$
and $X_\chi$, namely \bea X^R_u&=&-\fr 1 6 (5X_\eta+4X_\chi),\\
X^R_d&=&\fr 1 6 (7X_\eta-4X_\chi),\\ X^R_s&=&\fr 1 6
(X_\eta+8X_\chi),\\ X^R_{s_3}&=&\fr 1 6
(X_\eta-10X_\chi).\label{eq414} \eea The equations
(\ref{eq49})-(\ref{eq414}) are the fixing conditions for the
X-charges of the quark reps, therefore the charges of all fermion
reps are indeed fixed.

Knowing the $X$-charges of the multiplets, we get the electric
charges of their members as follows
 \bea q_u&=& \fr{5-\sqrt{3}\beta}{12},
\mbox{ }u=u,\mbox{ }c,\mbox{ }t,\\
q_d&=&-\fr{7+\sqrt{3}\beta}{12},\mbox{ } d=d,\mbox{ }s,\mbox{
}b,\\
q_{s_3}&=&-\fr{1+7\sqrt{3}\beta}{12},\\
q_s &=&-\fr{1-5\sqrt{3}\beta}{12},\mbox{ }s=s_1,\mbox{ }s_2.\eea
In addition, it is easy to check that all the remaining anomaly
cancelation conditions (\ref{a1}), (\ref{a4}) and (\ref{a5}) are
satisfied. So, the ECQ of the quark sector is also given with the
help of the anomaly cancelation.

We have the following remarks: \ben \item  We can check the
electric charge of the proton composed of three quarks $uud$ \bea
q_p&=&2q_u+q_d\crn &=&\fr{1-\sqrt{3}\beta}{4},\eea which yields
\be q_p=-q_e.\ee
\item The neutron is composed of three quarks $ddu$; therefore,
its electric charge is given by\bea q_n&=&q_u+2q_d\crn
&=&-\fr{3+\sqrt{3}\beta}{4},\eea which yields the following
interesting consequence \be q_n=-q_{\nu}.\ee
\een

As mentioned above, the coefficient $\beta$ should be fixed from
the known hypercharge values in the SM \cite{chli}. Hence, for a
lepton triplet, we have
 \bea Y(3_l)&=&\left(-\fr 1 2, -\fr 1 2,
+1\right)^T=\beta T_8+X^L_l.\label{beta}\eea From (\ref{eqe}) and
(\ref{beta}), it  follows that $\beta=-\sqrt{3}$. Then the
electric charges get the correct values as follows \bea
q_{\nu_e}&=&0,\mbox{ }
e=e,\mbox{ }\mu,\mbox{ }\tau,\\
q_{e}&=&-1,\mbox{ }
e=e,\mbox{ }\mu,\mbox{ }\tau,\\
q_u&=& +\fr 2 3,
\mbox{ }u=u,\mbox{ }c,\mbox{ }t,\\
q_d&=&-\fr 1 3,\mbox{ } d=d,\mbox{ }s,\mbox{
}b,\\
q_{s_3}&=&+\fr 5 3,\\
q_s &=&-\fr 4 3,\mbox{ }s=s_1,\mbox{ }s_2.\eea

These relations have also been  found in the literature
\cite{prs}, but they are based on the two principal conditions
such as the classical constraints (to generate mass for the all
fermions) and the anomaly cancelation.

\subsection{The ECQ in the 3-3-1  model with RH neutrinos}

For  the 3-3-1 model with RH neutrinos, the $\beta $  takes a
value of $-\fr{1}{\sqrt{3}}$. Therefore, the exotic quarks get the
electric charges different from those in the minimal model as
follows \bea q_{s_3}&=&+\fr 2 3,\\ q_s &=&-\fr 1 3,\mbox{
}s=s_1,\mbox{ }s_2.\eea This means that this model does not
contain the exotic charges.

The electric charges of the usual leptons and quarks are the same
as in the minimal model.

Thus, basing on the parity invariance of the electromagnetic
interaction and the anomaly cancelation,  we have  shown that the
usual 3-3-1 models contain in their framework the quantization of
the electric charge.
\section{Conclusions}
\label{sec:conclusions}

Analyzing the photon eigenstate structure, we have shown that the
general properties of the electromagnetic interaction such as the
parity invariance is properly kept in the framework of the 3-3-1
models. This is  a natural consequence of the conservation of the
electric charge. As a result, the electric charge quantization in
the usual 3-3-1 models has been derived.

Examining the fermion contents, we have found that the 3-3-1
models contain themselves two solutions such as the electric
charge quantization and the generation number problem. Moreover,
theoretically, the electric charges of the neutron and of the
neutrino as well as of the proton and of the electron are
opposite.

We pointed out that the electric charge quantization is
independent on generating mass to the fermions. This is the main
difference between our approach and that in \cite{prs}, which was
based on.

We have also shown that if in the  GUT, the electric charge
quantization results from  the algebra structure, here in the
3-3-1 models  based on the semi-simple group, it is  a direct
consequence of  the fermion contents.

This conclusion adds one more nice feature to the 3-3-1 models.

\section*{Acknowledgments}

This work was supported in part by National Council for
Natural Sciences of Vietnam contract No: KT - 41064.\\[0.3cm]


\end{document}